\documentclass[twocolumn,showkeys,showpacs,preprintnumbers,amsmath,amssymb,nofootinbib,floatfix]{revtex4}

\usepackage[dvips]{graphicx}
\usepackage{dcolumn}
\usepackage{bm}

\begin{document}

\title{Composite Higgs Boson in the Unified Subquark Model of All Fundamental Particles and Forces}

\author{Hidezumi Terazawa}
\email{terazawa@mrj.biglobe.ne.jp}
\affiliation{\vspace{3mm}Center of Asia and Oceania for Science(CAOS), 3-11-26 Maesawa, Higashi-kurume, Tokyo 203-0032, Japan\\}
\affiliation{Midlands Academy of Business \& Technology(MABT), Mansion House, 41 Guildhall Lane, Leicester LE1 5FR, United Kingdom}
\author{\vspace{1mm}Masaki Yasu\`{e}}%
\email{yasue@keyaki.cc.u-tokai.ac.jp}
\affiliation{\vspace{3mm}%
\sl Department of Physics, Tokai University,\\
4-1-1 Kitakaname, Hiratsuka, Kanagawa 259-1292, Japan\\
}

\date{January, 2014}

\begin{abstract}
In the unified subquark model of all fundamental particles and forces, the mass of the Higgs boson in the standard model 
of electroweak interactions ($m_H$) is predicted to be about $2\sqrt{6}m_W/3$ (where $m_W$ is the mass of the charged 
weak boson), which agrees well with the experimental values of $125-126$ GeV recently found by the ATLAS and CMS 
Colaborations at the LHC. It seems to indicate that the Higgs boson is a composite of the iso-doublet subquark-antisubquark 
pairs well described by the unified subquark model with either one of subquark masses vanishing or being very small 
compared to the other.
\end{abstract}


\maketitle
What most of us could expect to find in high energy experiments at the Large Hadron Collider was the Higgs boson ($H$), 
which was the only fundamental particle that had not been found in the standard model of electroweak interactions \cite{StandardModel}. 
In the unified composite models of all fundamental particles and forces \cite{SubquarkModelReview}, 
where not only quarks and leptons but also gauge bosons as well as the Higgs boson are all composites of subquarks, the more fundamental particles in nature, 
the mass of the Higgs boson has been predicted 
in the following three ways: 

In general, in composite models of the Nambu-Jona-Lasinio type \cite{NJLModel}, the Higgs boson appears as a composite state of fermion-
antifermion pairs with the mass twice as much as the fermion mass. The unified subquark model of the Nambu-Jona-Lasinio type
\cite{UnifiedSubquarkModel} has predicted the following two sum rules: 
\begin{equation}
m_W=[3(m_{w_1}^2+m_{w_2}^2)/2]^{1/2}
\end{equation}
and 
\begin{equation}
m_H=2[(m_{w_1}^4+m_{w_2}^4)/(m_{w_1}^2+m_{w_2}^2)]^{1/2}, 
\end{equation}
where $m_{w_1}$ and $m_{w_2}$ are the masses of the weak-iso-doublet spinor subquarks called \lq\lq wakems\rq\rq\ standing 
for weak and electromagnetic ($w_i$ for $i=1,2$) while $m_W$ and $m_H$ are the masses of the charged weak boson ($W$) and 
physical Higgs boson in the standard model, respectively. By combining these sum rules, the following relation has been 
obtained 
if the subquarks are iso-symmetric as 
$m_{w_1}=m_{w_2}$: 
\begin{equation}
m_w:m_W:m_H=1:\sqrt{3}:2. 
\end{equation}
From this relation, the wakem and Higgs boson masses have been predicted as 
\begin{equation}
m_w=m_W/\sqrt{3}=46.4 ~{\rm GeV} 
\end{equation}
and 
\begin{equation}
m_H=2m_W/\sqrt{3}=92.8 ~{\rm GeV} 
\end{equation}
for $m_W=80.4$ GeV \cite{ParticleDataGroup}.
On the other hand,if $m_{w_1}=0$ or $m_{w_2}=0$, the other relation can be obtained: 
\begin{equation}
m_w:m_W:m_H=1:\sqrt{3/2}:2. 
\end{equation}
From this relation, the non-vanishing wakem and Higgs boson masses can be predicted as 
\begin{equation}
m_w=m_W/\sqrt{3/2}=65.6 ~{\rm GeV} 
\end{equation}
and 
\begin{equation}
m_H=2m_W/\sqrt{3/2}=131 ~{\rm GeV} 
\end{equation}
for $m_W=80.4$ GeV \cite{ParticleDataGroup}.
More generally, from the two sum rules, the Higgs boson mass can be bounded as 
\begin{equation}
92.8 ~{\rm GeV}=2m_W/\sqrt{3}\leq m_H\leq 2\sqrt{6}m_W/3=131 ~{\rm GeV}. 
\\
\end{equation}

In the unified quark-lepton model of the Nambu-Jona-Lasinio type \cite{UnifiedSubquarkModel}, the following two sum rules for $m_W$ and $m_H$ have 
been predicted: 
\begin{equation}
m_W=(3<m_{q,l}^2>)^{1/2} 
\end{equation}
and 
\begin{equation}
m_H=2(\sum m_{q,l}^4/\sum m_{q,l}^2)^{1/2}, 
\end{equation}
where $m_{q,l}$\rq s are the quark and lepton masses and $<>$ denotes the average value for all the quarks and leptons. If 
there exist only three generations of quarks and leptons, these sum rules completely determine the top quark and Higgs 
boson masses \cite{SubquarkModel_mt_mH} as 
\begin{equation}
m_t\cong (2\sqrt{6}/3)m_W=131 ~{\rm GeV} 
\end{equation}
and 
\begin{equation}
m_H\cong 2m_t\cong (4\sqrt{6}/3)m_W=263 ~{\rm GeV}. 
\end{equation}

Furthermore, triplicity of hadrons, quarks, and subquarks \cite{Triplicity} tells us that these sum rules can be further extended to the 
approximate sum rules of 
\begin{equation}
m_W\cong (3<m_{B,l}^2>)^{1/2} 
\end{equation}
and 
\begin{equation}
m_H\cong 2(\sum m_{B,l}^4/\sum m_{B,l}^2)^{1/2}, 
\end{equation}
where $m_{B,l}$s are the \lq\lq canonical baryon\rq\rq\ and lepton masses and $<>$ denotes the average value for all the 
canonical baryons and leptons. The \lq\lq canonical baryon\rq\rq\ means either one of $p,n$ and other ground-state
 baryons of spin 1/2 and weak-isospin 1/2 consisting of a quark heavier than the $u$ and $d$ quarks and a scalar and 
isoscalar diquark made of $u$ and $d$ quarks. If there exist only three generations of quarks and leptons, these sum rules 
completely determine the masses of the canonical topped baryon, $T$, and the Higgs boson as 
\begin{equation}
m_T\cong 2m_W=161 ~{\rm GeV} 
\end{equation}
and 
\begin{equation}
m_H\cong 2m_T\cong 4m_W=322 ~{\rm GeV}. 
\end{equation}

Therefore, if the Higgs boson is found with the mass between 92.8 GeV and 131 GeV, it looks like a composite state of 
subquark-antisubquark pairs. If it were found heavier with $m_H$ around 263 GeV or even 322 GeV, it could be taken as a bound 
state of $t\overline{t}$ (\lq\lq topponium\rq\rq) or $T\overline{T}$ (\lq\lq topped-baryonium\rq\rq), respectively. If it 
were found with the mass lying between these typical masses, it might be taken as a mixture of subquark-antisubquark pairs 
and quark-antiquark pairs, \textit{etc.}.

Very recently, the ATLAS and CMS Collaboration experiments at the CERN Large Hadron Collider have almost excluded the two 
ranges for the Higgs boson mass: the one lower than 114 GeV and the other between 141 GeV and 476 GeV \cite{ATLAS,CMS}, which disagrees 
with both the prediction in the unified quark-lepton model of the Nambu-Jona-Lasinio type \cite{UnifiedSubquarkModel} and that in the unified 
baryon-lepton model of the Nambu-Jona-Lasinio type \cite{Triplicity}. Instead, the prediction in the unified subquark model \cite{UnifiedSubquarkModel} ($92.8 ~{\rm GeV}\leq m_H\leq 131 ~{\rm GeV}$) shows a right ballpark on which the mass of the Higgs boson in the standard model should land. 
Moreover, the fact that the experimental values of $m_H = 125-126$ GeV recently found by the ATLAS and CMS Collaborations 
are very close to the predicted one of $m_H = 2\sqrt{6}m_W/3 = 131 ~{\rm GeV}$ seems to strongly suggest that either one of 
$m_{w_1}$ and $m_{w_2}$ vanishes or is much smaller than the other. In fact, if indeed $m_H = 126$ GeV \cite{ATLAS,CMS}, the two sum 
rules completely determine the subquark masses in the unified subquark model as $(m_{w_1}, m_{w_2}) = (13.3 ~{\rm GeV}, 64.3 ~{\rm GeV})$ 
or $(64.3 ~{\rm GeV}, 13.3 ~{\rm GeV})$ for $m_W = 80.4 ~{\rm GeV}$ \cite{ParticleDataGroup}. It seems to indicate that the Higgs boson is a composite of the isodoublet 
spinor subquark-antisubquark pairs well described by the unified subquark model with either one of subquark masses vanishing 
or being very small compared to the other. 
We believe that this conclusion of the present paper is not only very
important in high energy physics but also very intriguing in physics or in
science in general.
Let us hope that the future LHC (and also ILC) experiments will tell us whether 
the unified subquark model is a viable model of all fundamental particles and forces!

\vspace{5mm}
\textbf{\large Acknowledgements}

\vspace{5mm}
 One of the authors (H.T.) thanks Professor Keiichi Akama and Professor Yuichi Chikashige for very useful helps in correcting errors in the 
original manuscripts.



\end{document}